\documentclass[11pt,fleqn,a4paper]{article}

\usepackage{latexsym,upgreek}
\usepackage{amssymb,amsfonts}
\usepackage[dvips]{graphicx}
\usepackage{cite}
\usepackage{euscript} 
\usepackage{bm,slashed}        
\usepackage{amsmath}   
\usepackage{mathrsfs}
\catcode`\@=11
\@addtoreset{equation}{section}
\def\theequation{\arabic{section}.\arabic{equation}}

\def\appendix{\renewcommand{\thesection}{\Alph{section}}\setcounter{section}{0}
              \renewcommand{\theequation}
            {\mbox{\Alph{section}.\arabic{equation}}}\setcounter{equation}{0}}

\newcommand{\pacs}[1]{\smallskip\noindent{\sl PACS numbers:
                       \hspace{0.3cm}#1}\par\bigskip\rm}
\def\babs{\hrule\par\begin{description}\item{Abstract: }\it} 
\def\eabs{\par\end{description}\hrule\par\medskip\rm}

\def\ea{\end{eqnarray}}      
\def\ba{\begin{eqnarray}}
\def\be{\begin{equation}}
\def\eeq{\end{equation}}

\def\vv{\upsilon}

\def\R{{\hbox{{\rm I}\kern-.2em\hbox{\rm R}}}}   
\def\H{{\hbox{{\rm I}\kern-.2em\hbox{\rm H}}}}   
\def\N{{\hbox{{\rm I}\kern-.2em\hbox{\rm N}}}}   
\def\C{{\ \hbox{{\rm I}\kern-.6em\hbox{\bf C}}}} 
\def\Z{{\hbox{{\rm Z}\kern-.4em\hbox{\rm Z}}}}   

\def\al{\alpha}

\begin{document}

\author{R.~Di Criscienzo\thanks{rdicris@physics.utoronto.ca}\\
Department of Physics-University of Toronto,\\
60 St. George Street, Toronto, Ontario, M5S 1A7, Canada\\
L.~Vanzo\thanks{vanzo@science.unitn.it}\\
Dipartimento di Fisica, Universit\`a di Trento \\
and INFN, Gruppo Collegato di Trento\\
Via Sommarive 14, 38100 Povo (TN), Italia} 
\date{}

\title{Fermion Tunneling from Dynamical Horizons}

\maketitle


\abstract{The instability against emission of fermionic particles by
  the  trapping horizon of an evolving black hole is analyzed using
  the Hamilton-Jacobi tunneling method. This method automatically
  selects one 
  special expression for the surface gravity of a changing horizon. 
The results also apply to point masses embedded in an expanding
  universe. As a bonus of the tunneling method, we gain the insight
  that the surface gravity still defines a temperature parameter as
  long as the evolution is sufficiently slow that the black hole pass
  through a sequence of quasi-equilibrium states, and that black holes
  should be semi-classically unstable even in a hypothetical world
  without bosonic fields.}  

\pacs{04.70.-s,04.70.Bw,04.70.Dy}

\section{Introduction}

In a previous paper\cite{DiCriscienzo:2007fm} we considered the
quantum instability of dynamical black holes using a
variant of the tunneling method introduced by Parikh and
Wilczek\cite{parikh}, according to which the Hawking
effect can be considered as a kind of tunneling transition
through the horizon of the black hole (BH). Probably, for such matters 
it was first 
applied in condensed matter physics, specifically to exotic phenomena
involving the analogue of horizons in superfluid $^3He$ films
\cite{Volovik:1999fc}. 
The method was refined and extended to more general cases in
\cite{svmp} and 
others papers as well \cite{others}, including the back
reaction effects~\cite{obr} and its extension to non-commutative
Schwarzschild space-time~\cite{Banerjee:2008gc}. For criticism and
counter 
criticism see also \cite{anm}, and \cite{Pilling:2007cn} for the
relation between tunneling and thermodynamics.\\
The Hawking's effect should not be confused with another tunneling
process, the escape of the particle to infinity through the
potential barrier surrounding the horizon, and about which the
Parikh-Wilczek method has nothing to say. \\
The variant mentioned above is the Hamilton-Jacobi 
method introduced in\cite{Visser:2001kq,angh:2005}, so
called after the  
appearance of a complete comparison analysis with the
Parikh-Wilczek method done in\cite{mann}. The tunneling
method provides not only new physical insight to an understanding of
the black hole radiation, but is also a powerful way to
compute the surface gravity for a vast range of solutions. Now for non
stationary black holes, termed dynamical black holes
in\cite{DiCriscienzo:2007fm}, things are not so simple and even the
possibility of Hawking radiation is in principle questionable, since
in general a changing horizon is not a null hypersurface, although it
is still one of infinite red shift. In particular, several definitions
of the surface gravity for evolving horizons have been proposed in the
past, all fitting the first law of BH mechanics more or less equally
well. A comparison is discussed throughly
in\cite{Nielsen:2007ac}. Among these the one proposed implicitly by
Visser \cite{Visser:2001kq} and more explicitly by Hayward and
Kodama\cite{Hayward:1997jp,Kodama:1979vn} is the most interesting to
us, as it is the one the tunneling method leads to. Still different,
otherwise reasonable, resuls are advocated for expanding
cosmological black holes \cite{Faraoni:2007gq,Saida:2007ru}, but we
shall not dwell about them here.\\
As we said above, a way to understand Hawking radiation is by means of
tunneling of particles through black-hole horizons. Such tunneling
approach uses the fact that the WKB 
approximation of the tunneling probability for the classical forbidden
trajectory from inside to outside the horizon is: 
\begin{equation}
\Upgamma \propto e^{- \frac{2}{\hbar}\mbox{ Im } I},
\label{prob} 
\eeq  
where $I$ is the classical action of the trajectory, to leading order
in $\hbar$. 
What kind of particles do we expect to find in the Hawking radiation
spectrum of a black-hole? In principle, all the Standard Model
particles. However, most of the calculations in literature have been
performed just for scalar fields, except in~\cite{Kerner:2007rr},
where a detailed study of spin one-half emission was considered for
stationary black holes, and~\cite{Li:2008ws} for the special case of
the BTZ black hole.\\     
What we are going to check is that the tunneling approach via the
Hamilton-Jacobi method is consistent with the Kodama-Hayward
prediction even for $1/2-$spin particles in dynamical black holes and
that, as a consequence, they should be semi-classically unstable even
in a world with no bosonic fields.

\section{Bardeen-Vaidya space-times and cosmological black holes}

Let us consider the Bardeen - Vaidya metric (BV) in $D=4$
dimensions~\cite{vaidya,bardeen}

\begin{equation}
ds^2=-e^{2\psi(r,v)}A(r,v)dv^2+2 e^{\psi(r,v)}dv dr+
r^2d\omega^2,
\label{efbard}
\end{equation}
where $v$ is an advanced time null coordinate, $d\omega^2=d\theta^2 +
\sin^2\theta\, d\phi^2$ and, in the simplest
case we have in mind, $A(r,v)$ is  just 
\begin{equation}
A(r,v)= 1 - \frac{2m(r,v)}{r}.
\end{equation}
The inverse metric is given by:
\begin{equation}
g^{\mu\nu} = \left(\begin{array}{cccc}
0 & e^{-\psi(r,v)} & 0 & 0\\
e^{-\psi(r,v)} & A(r,v) & 0 &0\\
0& 0& \frac{1}{r^2} &0\\
0&0&0&\frac{1}{r^2 \sin^2\theta}
\end{array}\right). \label{invmetric}
\end{equation}
This class of metrics possesses - under
very general conditions - a trapping horizon (TH), as defined by
Hayward\cite{Hayward:1997jp}, which in the present case is given by
the equation $A(r,v)=0$. This defines a curve $r=r_H(v)$ giving the
location of the apparent horizon; the quantity $E=m(r_H(v),v)$ is the
Misner-Sharp mass\cite{hawk} of the horizon, in term of which the
horizon will be trapping if $m^{'}(r_H,v)<1/2$, a prime denoting the
radial derivative. The TH proved to be the key concept in such matters
as particles emission and gravitational entropy (for a very recent
discussion see~\cite{Nielsen:2008dj}), and can be applied successfully
even to higher dimensional Vaidya space-times~\cite{Ren:2007xw}. \\ 
According to the Kodama-Hayward theory,
to such TH is associated a geometrical surface gravity 
\begin{equation}
\kappa(v) =
\frac{A'(r,v)}{2}_{\big\vert_{TH}}=\frac{1}{2r_H}-\frac{m^{'}(r_H,v)}{r_H}
\label{KH}\,. 
\end{equation}
We see the meaning of the trapping condition: it ensures the
positivity of the surface gravity.\\
The second example we are interested in is the McVittie solution
\cite{mcv} for a point mass in a Friedmann-Robertson-Walker flat
cosmology. In isotropic spatial coordinates it is given by\cite{mcv} 
\be
ds^2=-A(\rho,t)dt^2+B(\rho,t)\left( d\rho^2+\rho^2d\omega^2 \right)\,
\eeq
with
\be
A(\rho,t)=\left[ \frac{1-\frac{m}{a(t)\rho}}{1+
\frac{m}{a(t)\rho}} \right]^2\,, \qquad
 B(\rho,t)=a(t)^2 \left[ 1-\frac{m}{a(t)\rho}\right]^{2}\nonumber
\eeq
When the mass parameter $m=0$, it reduces to a spatially flat FRW
solution with scale factor $a(t)$; when $a(t)=1$ it reduces to the
Schwarzschild metric with mass $m$.  
In four dimensions this solution has had a strong impact on the
general problem of matching the Schwarzschild solution with cosmology,
a problem faced also by Einstein and Dirac. 
Besides McVittie, it has been extensively studied by Nolan 
in a series of papers \cite{nolan}. To put the metric in the general
form of Kodama theory, we use what may be called the Nolan gauge, in
which  the metric reads 
\begin{eqnarray}\label{nolan}
ds^2 = -\left(A_s-H^2(t)r^2\right) dt^2+A_s^{-1}dr^2 
- 2A_s^{-1/2}H(t)r\,drdt+r^2d\omega^2
\end{eqnarray}
where $H(t)=\dot{a}/a$ is the Hubble parameter and, for example, in
the charged case, $A_s=1-2m/r+q^2/r^2$. 
 In passing to the Nolan gauge a choice of sign in the cross term
$drdt$ has been done, corresponding to an expanding universe; the
transformation $H(t)\to-H(t)$ changes this into a contracting one. 
In the following we shall consider $q=0$; then the
Einstein-Friedmann equations  read
\be\label{eeqs}
3H^2=8\pi\rho\,, \qquad 2A_s^{-1/2}\dot H(t)+3H^2=-8\pi p\,.
\eeq
It follows that $A_s=0$, or $r=2m$, is a curvature singularity similar
to $r=0$ in FRW models,  namely it is a big bang singularity.  
When $H=0$ one has the Schwarzschild solution. The term
$H^2r^2$ in the metric strongly resembles a varying cosmological
constant; in fact if $H$ is constant the metric reduces to the
Schwarzschild-de Sitter solution in Painlev\'e coordinates. 
The trapping horizon is the root of $A_s=H^2r^2_H$, and is time
dependent. The Misner-Sharp mass and the geometrical surface gravity
are, respectively,   
\be\label{ms}
E=m+\frac{1}{2}\,H(t)^2r_H^3\,,
\eeq
\be
\kappa(t)=\frac{m}{r_H^2}-H^2 r_H -\frac{ \dot H}{2 H}\,.
\label{sg}
\eeq
Note that $E=r_H/2$. In the static cases everything agrees with the
standard  results. The 
 surface gravity has an interesting expression in terms of the sources
of Einstein equations and the Misner-Sharp mass. Let $T_{2D}$ be the
reduced trace of the stress tensor in the space normal to the sphere
of symmetry, evaluated on the horizon. For the Vaidya-Bardeen
metric it is, by Einstein's equations,
\be
T_{2D}=T^v_{\;v}+T^r_{\;r}=-\frac{1}{2\pi r_H}\,\frac{\partial
  m}{\partial r}_{|r=r_H}
\eeq
For the McVittie's solution, this time by Fredmann's equations
\eqref{eeqs} one has  
\be
T_{2D}=-\rho+p=-\frac{1}{4\pi}\left( 3H^2+\frac{\dot{H}}{Hr_H}\right)
\eeq 
We have then
\be
\kappa=\frac{E}{r_H^2}+2\pi r_HT_{2D}\,.
\eeq 
It is worth mentioning the pure FRW case, i.e. $A_s=1$, for which
$\kappa(t)= -H(t)-\dot H/2 H$.
We feel that these expressions for the surface gravity are non trivial
and display deep connections with the emission process. We now proceed to
analyze the Dirac equation in BV space-time. Since the Hawking's
effect is a phase phenomenon, which is why the tunneling method
works so well, we expect the spinor amplitudes to play a minor
role. We verify this for the BV ``shining star'', but use a more
straightforward method for the McVittie's solution. 

\section{Dirac equation for BV metric} The Dirac equation is,
\be
\left(\gamma^\mu D_\mu + \frac{m}{\hbar}\right)
\Uppsi(v,r,\theta,\phi) = 0, \label{Dirac} 
\eeq
where
\ba
D_\mu &=& \partial_\mu + \frac{i}{2} {\Gamma^\alpha_{\;\mu}}^{\;\;\beta} \,
\Sigma_{\alpha\beta} , \label{diffop} \\ 
\Sigma_{\alpha\beta} &=& \frac{i}{4} [\gamma_\alpha,
  \gamma_\beta]_-.\label{Comm} 
\ea
The $\gamma^\mu -$ matrices satisfy the Clifford algebra,
\be
[\gamma_\alpha,\gamma_\beta]_+ = 2 g_{\alpha\beta} \mathbb{I},
\eeq
where $\mathbb{I}$ is the $(4\times 4)$-identity matrix.
In order to get the Dirac-$\gamma$ matrices for the BV metric at hand,
we define firstly a tetrad of orthogonal vectors $E^a_\mu$ s.t. 
\be
\eta_{ab}E^a_\mu E^b_\nu = g_{\mu\nu}.
\eeq   
The convention is that (first) latin indices are Minkovskian so they
run over $(0,1,2,3)$; greek indices are coordinate indices so they run
over $(v,r,\theta,\phi)$. \\ 
Of course there are many different tetrads, but the simplest choice is
the following: 
\ba
E^a_v &=& \left(e^\psi \sqrt{A}, 0,0,0\right) , \label{tetrad1}\\
E^a_r &=&\left(-\frac{1}{\sqrt{A}}, \frac{1}{\sqrt{A}},0,0\right),\\
E^a_\theta &=& \left(0,0,r,0\right),\\
E^a_\phi &=& \left(0,0,0,r\sin\theta\right).
\label{tetrad4}
\ea
The $\gamma_\mu -$matrices are expressed in terms of the tetrad in the
following way: 
\be
\gamma_\mu = \gamma_a E^a_\mu .
\eeq
With our choice (\ref{tetrad1}) - (\ref{tetrad4}), it turns out that
\ba
\gamma_v &=& e^\psi \sqrt{A} \, \gamma_0 ,\\
\gamma_r &=& \frac{1}{\sqrt{A}} \left(\gamma_1 - \gamma_0\right),\\
\gamma_\theta &=& r\,\gamma_2 ,\\
\gamma_\phi &=& r\sin\theta \,\gamma_3 ;
\ea
and
\ba
\gamma^v &=& \frac{e^{-\psi}}{\sqrt{A}} \left(\gamma_1 - \gamma_0\right) ,\\
\gamma^r &=& \sqrt{A} \,\gamma_1,\\
\gamma^\theta &=& \frac{1}{r}\,\gamma_2 ,\\
\gamma^\phi &=& \frac{1}{r\sin\theta} \,\gamma_3 .
\ea
We can also express the
\be
\gamma^5 \stackrel{def}{=} i
\gamma^v\,\gamma^r\,\gamma^\theta\,\gamma^\phi = \frac{i
  e^{-\psi}}{r^2 \sin\theta} \left(\mathbb I - \gamma_0
\gamma_1\right) \gamma_2\gamma_3. 
\eeq
All what we need now is a convenient representation of the
Dirac-$\gamma_a$ matrices satisfying $[\gamma_a,\gamma_b]_+ = 2
\eta_{ab}$ with $\eta = \mbox{diag}(-1,1,1,1)$. 
\ba
\gamma_0 &=& \left(\begin{array}{cc}
i & 0 \\
0 & -i
\end{array}\right);  \\
\gamma_1 &=& \left(\begin{array}{cc}
0 & \sigma_3 \\
\sigma_3 & 0
\end{array}\right);  \\
\gamma_2 &=& \left(\begin{array}{cc}
0 & \sigma_1 \\
\sigma_1 & 0
\end{array}\right);  \\
\gamma_3 &=& \left(\begin{array}{cc}
0 & \sigma_2 \\
\sigma_2 & 0
\end{array}\right).
\ea
The $\sigma$-matrices are the Pauli matrices satisfying the usual
relations: 
\be
\sigma_i \sigma_j = \mathbb{I}^{(2\times 2)} \delta_{ij} + i
\varepsilon_{ijk}\sigma_k ,\qquad i,j,k =1,2,3\;. 
\eeq
By virtue of (\ref{efbard}) and
(\ref{invmetric}), one calculates the
${\Gamma^\alpha_{\;\mu}}^{\;\;\beta}$ symbols required in
(\ref{diffop}). We list the result in the footnote below\footnote{ The
  non-vanishing ${\Gamma^\al_{\;\mu}}^{\;\;\beta} = g^{\beta\nu}
  \Gamma^\al_{\;\mu\nu}$ are: 
\ba
{\Gamma^\vv_{\;\vv}}^{\;\;\vv} &=& e^{-\psi} \dot\psi + \frac{A'}{2} +
A\psi' \;, \qquad {\Gamma^\vv_{\;\theta}}^{\;\;\theta} =
-\frac{e^{-\psi}}{r}\;, \qquad {\Gamma^\vv_{\;\phi}}^{\;\;\phi} =
-\frac{e^{-\psi}}{r}\; ;\nonumber \\ 
{\Gamma^r_{\;\vv}}^{\;\;\vv} &=& -\frac{A'}{2} - A\psi' \;, \qquad
{\Gamma^r_{\;\vv}}^{\;\;r} = - \frac{\dot A}{2}  \;, \qquad
{\Gamma^r_{\;r}}^{\;\;r} = - \frac{A'}{2}\;,
{\Gamma^r_{\;r}}^{\;\;\vv} = e^{-\psi} \psi'\;,
\qquad{\Gamma^r_{\;\theta}}^{\;\;\theta} = - \frac{A}{r} \;, \qquad
      {\Gamma^r_{\;\phi}}^{\;\;\phi} =  - \frac{A}{r}\; ; \nonumber \\ 
{\Gamma^\theta_{\;r}}^{\;\;\theta} &=& \frac{1}{r^3}\;, \qquad
{\Gamma^\theta_{\;\phi}}^{\;\;\phi} = -
\frac{\cos\theta}{r^2\sin\theta} \;,\qquad
     {\Gamma^\phi_{\;\theta}}^{\;\;\phi} =
     \frac{\cos\theta}{r^2\sin^3\theta} \; ;\nonumber \\ 
{\Gamma^\phi_{\;\phi}}^{\;\;\theta} &=&
\frac{\cos\theta}{r^2\sin\theta} \;, \qquad
     {\Gamma^\phi_{\;r}}^{\;\;\phi} =  \frac{1}{r^3\sin^2\theta}
     \nonumber . 
\ea 
}
for the sake of completness. Next, we should calculate the
$\Sigma_{\alpha\beta}$ defined in (\ref{Comm}): a task that Maple can do
very quickly; in the end, it remains to evaluate the combination of
such results, namely   
\be
\slashed{\partial} \Uppsi + i\left[\frac{1}{2} \gamma^\mu
  {\Gamma^\alpha_{\;\mu}}^{\;\;\beta} \, \Sigma_{\alpha\beta} \Uppsi \right]
+ \frac{m}{\hbar} \Uppsi = 0 . \label{useless} 
\eeq
Let us employ the following ansatz for the spin-up Dirac
field\footnote{We shall perform a detailed analysis only for the
  spin-up case, being confident that \textit{mutatis mutandis}
  everything applies in the same way to the spin-down case.}: 
\be
\Uppsi(\vv,r,\theta,\phi)_{\uparrow} = \left(\begin{array}{c}
\Xi(\vv,r,\theta,\phi) \\
0\\
\Omega(\vv,r,\theta,\phi)\\
0
\end{array}\right) \exp{\left[\frac{i}{\hbar}
  I_{\uparrow}(\vv,r,\theta,\phi)\right]}. \label{spinor} 
\eeq
Plugging the ansatz (\ref{spinor}) into Dirac equation
(\ref{useless}), it turns out that the term in square brackets is of
order $O(\hbar)$. Thus, we do not need to work out its precise form,
since in the $\hbar \rightarrow 0$ limit it vanishes.  
To leading order in $\hbar$ equation (\ref{useless}) becomes
$0= \left( \hbar \slashed{\partial} + m \right)  \Uppsi_{\uparrow} +
O(\hbar)$, or
\begin{eqnarray}
0 &=& \exp{\left[\frac{i}{\hbar}
    I_{\uparrow}(\vv,r,\theta,\phi)\right]}
\times \left\{ \frac{e^{-\psi}}{\sqrt{A}}  \left(\begin{array}{c} 
i \Xi \partial_\vv I_\uparrow - \Omega  \partial_\vv I_\uparrow \\
0\\
- \Xi \partial_\vv I_\uparrow -i \Omega \partial_\vv I_\uparrow \\
0
\end{array}\!\!\right)\right. \\
&&\left. + \sqrt{A} \left(\begin{array}{c}
-\Omega \partial_r I_\uparrow \\
0\\
- \Xi \partial_r I_\uparrow \\
0
\end{array}\right) + \frac{1}{r} \left(\begin{array}{c}
0 \\
- \Omega \partial_\theta I_\uparrow \\
0\\
- \Xi \partial_\theta I_\uparrow 
\end{array}\right) + \frac{1}{r\sin\theta} \left(\begin{array}{c}
0 \\
- i \Omega \partial_\phi I_\uparrow \\
0\\
- i \Xi \partial_\phi I_\uparrow 
\end{array}\right) + im \left(\begin{array}{c}
\Xi \\
0\\
\Omega\\
0
\end{array}\right) \right\} .\nonumber
\label{step2}
\end{eqnarray}
Thus, we get the following equations:
\ba
\vv &: & \quad \frac{i \Xi}{\sqrt{A}} \left(e^{-\psi} \partial_\vv\right)
I_\uparrow  -  \frac{ \Omega}{\sqrt{A}} \left(e^{-\psi} \partial_\vv\right)
I_\uparrow - \sqrt{A} \Omega \partial_r I_\uparrow + i\Xi m =0 ;
\label{veq}\\ 
r &: &\quad - \frac{\Omega}{r} \left( \partial_\theta I_\uparrow  +
\frac{i}{\sin\theta} \partial_\phi I_\uparrow\right) =0; \label{req}\\ 
\theta &:& \quad   -\frac{\Xi}{\sqrt{A}} \left(e^{-\psi} \partial_\vv
\right)I_\uparrow -  \frac{i \Omega}{\sqrt{A}} \left(e^{-\psi} \partial_\vv
\right)I_\uparrow - \sqrt{A} \Xi \partial_r I_\uparrow + i\Omega m =
0;\label{thetaeq}\\  
\phi &:& \quad - \frac{\Xi}{r} \left( \partial_\theta I_\uparrow  +
\frac{i}{\sin\theta} \partial_\phi I_\uparrow\right) =0,\label{phieq} 
\ea
and the Kodama vector $K = e^{-\psi} \partial_\vv$ has been put in evidence
throughout. $K$ plays the role of the Killing vector $\partial_t$ for
dynamical black-holes. Therefore, it makes sense the following ansatz
for the action: 
\be
I_\uparrow = - \int d\vv\, e^{\psi(r,\vv)} E + W(r) + J(\theta,\phi),
\eeq 
which inserted into (\ref{veq}) - (\ref{phieq}) gives:
\ba
\vv & : & \quad \frac{1}{\sqrt{A}}(i \Xi- \Omega) E - \sqrt{A} \Omega
W'(r) + i\Xi m =0 ; \label{veq2}\\ 
r & : &\quad - \frac{\Omega}{r} \left( J_\theta (\theta,\phi) +
\frac{i}{\sin\theta} J_\phi (\theta,\phi)\right) =0; \label{req2}\\ 
\theta &: & \quad  - \frac{1}{\sqrt{A}}(\Xi + i\Omega) E  - \sqrt{A}
\Xi W'(r) + i\Omega m = 0;\label{thetaeq2}\\ 
\phi &:& \quad - \frac{\Xi}{r} \left( J_\theta (\theta,\phi)  +
\frac{i}{\sin\theta} J_\phi (\theta,\phi)\right) =0 .\label{phieq2} 
\ea
(\ref{req2}) and (\ref{phieq2}) imply that $J(\theta,\phi)$ is a
complex function. The same solution for $J$ is
obtained for the spin-down case, then its contribution to the rate
emission $\Upgamma$ cancels out and we can forget about it.  
As regard the remaining equations, we have
\begin{enumerate}
\item $\Xi = i \Omega$, then
\be
W'(r) = - \frac{2 E}{A(r,\vv)} ;
\eeq
\item $\Xi =- i \Omega$, then
\be
W'(r) = 0 , \label{incoming} 
\eeq
\end{enumerate}
both in the massless and massive cases. Solution (\ref{incoming}) has
not to be considered surprising, since 
\be
\gamma^5 \Uppsi =  \left(\begin{array}{c}
i\Xi -\Omega \\
0\\
\Xi + i \Omega\\
0
\end{array}\right).
\eeq
meaning that case 2. corresponds to the incoming particle absorbed in
the classical limit
with probability $\mathscr{P}[incoming] =1$. The emission process is
described instead by case 1. As explained 
elsewhere~\cite{DiCriscienzo:2007fm} this implies that 
\be
\mbox{Im} W(r) = - \mbox{Im}\int d r\, \frac{E}{A(r,\vv) /2} = \frac{2\pi i
  E}{A'(r_{H}(\vv),\vv)} . 
\eeq 
In the end we get,
\be
\Upgamma \stackrel{(\ref{prob})}{\propto} e^{-
  \frac{2\pi}{A'(r_{H}(\vv),\vv)/2} E}, \quad
\Longrightarrow \quad T = \frac{A'(r,\vv)_{\vert_{r=r_H(\vv)}}}{4\pi}, 
\eeq
confirming the predictions of the Kodama-Hayward theory (\ref{KH}). We
conclude that our dynamical black hole is unstable against the
emission of spinor particles. It
is essential, for this result to hold, that the black hole be
slowly evolving on the timescale of the wave, since otherwise no
meaningful notion of a frequency is available.   

\section{MacVittie cosmological black hole}

The preceding calculations showed that the derivatives of the action
were strongly mixed by the matrix structure of the Dirac
equation. Nevertheless things combined so that only the radial
derivative of the action was really important. Hence we will not study
the full Dirac equation in the following, but use instead a
shortcut. Writing as before 
\be
\Uppsi=U\exp{\left[\frac{i}{\hbar}
  I(t,r,\theta,\phi)\right]}.  
\eeq 
where $U$ is a slowly varying spinor amplitude, from the Dirac
equation we get
\be
\slashed{D}U+\hbar^{-1}\left(i\slashed{\partial}I+m\right)U=0
\eeq
In the semi-classical limit the second term dominates, so
$\left(i\slashed{\partial}I+m\right)U=0$; thus the
matrix $\left(i\slashed{\partial}I+m\right)$ must be singular, or
equivalently,
\be
g^{\mu\nu}\partial_{\mu}I\partial_{\nu}I+m^2=0
\eeq
The action will have a simple pole at the location of the trapping
horizon affecting to outgoing modes, so for these we also neglect
the mass term. Then for an outgoing particle we find 
\be\label{outg}
\partial_rI=-F(r,t)^{-1}\partial_t I\,,
\eeq
where 
\be
F(r,t)=\sqrt{A_s(r)}(\sqrt{A_s(r)}-rH(t))\,.
\eeq 
We pick the imaginary part by expanding this function at the  horizon  
along a future null direction, using the fact that for two
neighbouring events on a null direction in the metric \eqref{nolan},
one has $t-t_0=(2H_0^2r_0^2)^{-1}(r-r_0)$, where $H_0=H(t_0)$. We find
the result 
\be\label{exp1}
F(r,t)\!=\!\left(\frac{1}{2}\,A^{'}_s(r_0)-r_0 H_0^2-\frac{\dot
  H_0}{2H_0}\right) (r-r_0)=\kappa_0(r-r_0)
\eeq
where this time $r_0=r_H(t_0)$.
From this equation we see that $\partial_rI$ has a simple pole at the
trapping horizon; hence, making use again of Feynman
$i\epsilon$-prescription, one finds 
\be
\mbox{Im}
I=\pi\kappa(t_0)^{-1}\omega(t_0), 
\eeq
where $\omega(t)=\partial_tI$ is the energy at time $t$,   
in complete agreement with the geometric evaluation of the previous
sections. However, we stress that a full justification of the given
shortcut really requires the full spinor amplitudes, as we showed
above while discussing the BV black hole.


\end{document}